\documentclass{epl}
\usepackage{amssymb}
\usepackage{amsmath}
\usepackage[all]{xy}

\def\idty{{\rm 1\mkern-5.4mu I}}

\begin{document}

\title{Comment on ``Exclusion of time in the theorem of Bell''\\
by K. Hess and W. Philipp}
\shorttitle{Exclusion of time in the theorem of Bell}
\author{R.D. Gill\inst{1}\thanks{E-mail: \email{gill@math.uu.nl}}
\and G. Weihs\inst{2,3}\thanks{E-mail: \email{gregor.weihs@stanford.edu}}
\and A. Zeilinger\inst{3}\thanks{E-mail: \email{anton.zeilinger@univie.ac.at}}
\and M. \.Zukowski\inst{4}\thanks{E-mail: \email{fizmz@univ.gda.pl}}
}
\shortauthor{R.D. Gill \etal}
\institute{
\inst{1}Mathematical Inst., Univ.~Utrecht - Budapestlaan 6, 3584 CD Utrecht, Netherlands\\
\inst{2}Ginzton Labs, Stanford Univ.\ - Stanford, California, 94305--4088, USA\\
\inst{3}Inst.~f\"ur~Experimentalphysik, Univ.~Vienna - Boltzmanngasse 5, 1090 Wien, Austria\\
\inst{4}Inst.\ Fizyki Teoret.~i~Astro. - Uniwersytet Gda\'nski, PL-80-952 Gda\'nsk, Poland
}
\pacs{03.65.Ud}{Entanglement and quantum nonlocality (e.g.\ EPR paradox,Ê 
Bell's inequalities, GHZ states, etc.)}

\maketitle

\begin{abstract}
A recent Letter by Hess and Philipp claims that Bell's theorem neglects
the possibility of time-like dependence in local hidden variables, hence
is not conclusive. Moreover the authors claim that they have constructed,
in an earlier paper,  a local realistic model of the EPR correlations. 
However, they themselves have neglected the experimenter's freedom to 
choose settings, while on the other hand, Bell's
theorem can be formulated to cope with time-like dependence. This 
in itself proves that their toy model cannot satisfy local realism, but
we also indicate where their proof of its local realistic nature fails.
{\tt Version: \today}
\end{abstract}

\section{Introduction}

The recent papers \cite{hp-quant-ph, hp-pnas1,hp-pnas2,hp-epl}
by Hess and Philipp have drawn a lot of attention, especially
since the work was featured on \emph{Nature}'s web pages and from there
reached the popular press in many countries.  
In this comment we would like to point out a number of fatal errors in the work,
of which the most serious is the lack of recognition of the choice which an 
experimenter is free to make in the laboratory, and which a theoretician 
is free to make in a \emph{Gedankenexperiment}. We shall convert this
\emph{freedom} into a statistical independence assumption, and show how it
plays a vital role in obtaining Bell's theorem: quantum mechanics
violates local realism.

In fact, Hess and Philipp criticize Bell's proof of the theorem 
on the grounds of a quite different independence assumption made by
Bell whose only role is to enable him to cover stochastic 
as well as deterministic
hidden variable models (his hidden variables $\lambda$
are not assumed to be localized, see \cite{bell} p.\ 153,
paragraph 4 of ``Bertlmann's socks and the nature of reality''). 
They claim that time variables should be included in the proof but as
we show below, this is untrue. 
Hess and Philipp's proof that their toy model of the EPR correlations satisfies 
local realism is incomplete (and uncompletable).
It is perhaps difficult to appreciate the assumptions and implications of
the theorem, but the issues which Hess and Philipp 
raise are all well known and have
been thoroughly discussed, without this leading to abandonment of the theorem,
for nearly forty years. The issue of time variation and time dependence is
interesting but irrelevant to the theorem.
It does not spoil experimental conclusions either, see \cite{gill}.
That paper takes care of another recent attempt, \cite{accardi1,accardi2,
accardi3}, to find fault with Bell's theorem. 

In order to be more specific about the errors in their results, we will first
present our own proof of Bell's theorem and discuss its assumptions,
emphasizing aspects of \emph{freedom} and \emph{control},
and then turn to a refutation of the arguments of Hess and Philipp.
What we say is not new. Our formulation is a summary of attempts
of many earlier papers to formulate very precisely the assumptions
behind the theorem of Bell.

\section{Freedom and control in Bell's theorem}

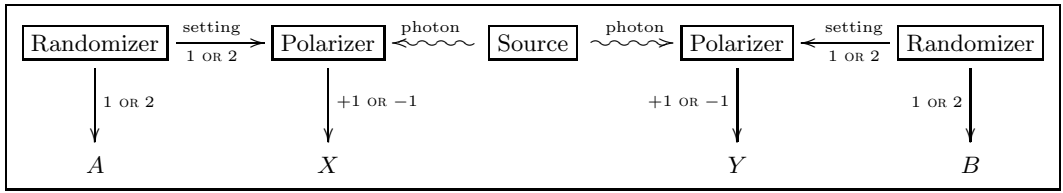
\begin{figure}
\begin{equation*}
\boxed{
\xymatrix@C+1em@R+1ex{
\boxed{\text{Randomizer}}\ar[d]^{1\text{\ \sc or\ }2} \ar[r]^{\text{setting}}_{1\text{\ \sc or\ }2}& 
\boxed{\text{Polarizer}} \ar[d]^{+1\text{\ \sc or\ }-1} & 
\boxed{\text{Source}} \ar@{~>}[l]_{\text{photon}} \ar@{~>}[r]^{\text{photon}} & 
\boxed{\text{Polarizer}} \ar[d]_{+1\text{\ \sc or\ }-1} & 
\boxed{\text{Randomizer}} \ar[l]_{\text{setting}}^{1\text{\ \sc or\ }2} \ar[d]_{1\text{\ \sc or\ }2}\\
   \overset{\  }A    &  \overset{\  }X  &    &   \overset{\  }Y  &   \overset{\  }B 
}
}
\end{equation*}
\caption{One trial of a Bell-type delayed choice experiment.}
\label{f.1}
\end{figure}

Figure~\ref{f.1} gives a schematic view of one trial (one pair of photons)
in a Bell-type delayed choice experiment, in particular, as in the experiment of
Weihs et al.\ \cite{weihsetal}, who for the first time fully implemented Bell's
requirement ``the filter settings are chosen during the flight of the 
photons''. See \cite{bell}, Figure 7, p.\ 151 
(paragraph 4 of ``Bertlmann's socks ...'').
We will use the words ``photons'', ``polarizer'' and so on, but of course the picture 
could be applied to many different physical realisations of the Bell singlet 
or other suitably entangled state.
In the two wings of the experiment, a ``setting'' is chosen by some random device. 
In each wing, there are two possible settings. We shall give the possible settings
the labels $1$, $2$. We let $A$ and $B$ denote the random setting actually chosen.
Thus $A$ and $B$ each take values say $i$, $j$ $\in\{1,2\}$.  The setting in each
wing is fed into a measurement device, just before a quantum particle arrives from
a distant source at the device. The measurement results in an outcome $\pm1$.
If only because the settings are random, the outcomes are (or could be) 
random too. We will denote the outcome left by $X$ and right by $Y$.

One should think of the two randomizers and the two polarizers
as being, all four, well separated from one another. To be concrete, 
consider the following rather elaborate randomization procedure: in the left wing 
of the experiment, Alice shuffles and cuts a pack of cards, and decides on the 
basis of the chosen card (red or black) how to encode a subsequent coin toss (H or T) as setting
$1$ or $2$, or vice versa. She selects a coin from her purse and tosses
it, using the encoding just determined in order to feed either a $1$ or a $2$
into the communication line to `her' polarizer. Far away, and simultaneously,
Bob follows a similar procedure.
Better still he uses other randomization devices such as a roulette wheel or
dartboard (he is a very poor darts player)
or pseudo-random number generator with seed chosen by tossing
dice. Note the freedom which the two persons have in: how many 
times to shuffle their pack of cards, which coin to pick from their purse, and so on.
And notice also the complexity of the path, which, even if one believes it is
essentially deterministic, results in the choice $1$ or $2$ at each polarizer.
We assume that the complete procedure used to generate 
$A$ and $B$ may be mathematically modelled as
\emph{independent, fair coin tosses}, thus: each of the four possible values $ij$ of 
the pair $AB$ are equally likely. In \cite{weihsetal} the randomizers were
actually systems based on quantum optics.

Now so far we have just introduced notation for the four variables 
$A$, $B$, $X$, $Y$ which actually get 
observed when the experiment (just one trial) is carried out. 
We now procede to 
describe what we mean by \emph{local realism} and show how this implies a
relationship between various probabilities concerning the observable variables.

The first element of local realism is \emph{realism} itself. By this we mean any
mathematical-physical model, or a scientific standpoint, which allows one to
introduce a further \emph{eight} variables into the the model so far, 
which we denote by $X_{ij}$, $Y_{ij}$, where $i,j=1,2$, and which are such that
\begin{equation}\label{e:realism}
X~\equiv~X_{AB},\quad Y~\equiv~Y_{AB}\quad\quad\mbox{(realism).}
\end{equation}
In words: one may conceive, as a thought experiment or as part of a mathematical
model, of ``what the measurement outcomes could be, under any of the possible
measurement settings''; you get to see the outcomes corresponding to the actually
selected settings. No hidden variables appear anywhere in our argument, beyond
these eight. However, given a (possibly stochastic) hidden variables theory, 
for instance that of \cite{hp-quant-ph, hp-pnas2},
one will be able to define our eight variables as
(possibly random) functions of the variables in that theory. These variables
coexist together independently of which experiment is \emph{actually} 
performed on either side.

Note that this assumption, by itself, can always be made: simply define all
$X_{ij}\equiv X$, and $Y_{ij}\equiv Y$!  However, that particular choice will be ruled out
by the next assumption but one (freedom). But first, we introduce
locality. The following is supposed to hold for all $i,j$:
\begin{equation}\label{e:locality}
X_{i1}~\equiv~X_{i2},\quad Y_{1j}~\equiv~Y_{2j}\quad\quad\mbox{(locality).}
\end{equation}
That is to say, the outcome which you would see left, under either setting, does
not depend on which setting might be chosen, right, and vice versa.
Working under the locality assumption, we
write
\begin{equation}\label{e:local}
X_i\equiv X_{ij},\quad Y_j\equiv Y_{ij}\quad\mbox{for each $i,j$}. 
\end{equation}

Finally we make one more assumption, which we call \emph{freedom}, 
often only tacit in treatments of Bell's theorem:
\begin{equation}\label{e:freedom}
(A,B)~~\mbox{is statistically independent of}~~(X_1,X_2,Y_1,Y_2)\qquad
\mbox{(freedom).}
\end{equation}
This \emph{freedom} assumption expresses that the choice of settings in the
two randomizers, summarized in the fair coin tosses $A$ and $B$,
is causally separated from the locally realistic mechanism which produces
the potential outcomes $X_1$, $X_2$; $Y_1$, $Y_2$; the actually observed
outcomes finally being selected as $X=X_A$, $Y=Y_B$. Statistical independence, or
complete lack of correlation, means that in many, many conceptual repetitions 
of the experiment, the relative frequencies with which the quadruple 
$(X_1,X_2,Y_1,Y_2)$ takes on any of its $2^4$ possible values, 
remain the same within each subensemble defined by each of the 
four possible values of $AB$ (and the other way round).

Contained in the above is an assumption of \emph{control}. When 
Alice and Bob send the chosen setting labels $i$, $j$ to their measurement devices,
they will likely cause some further unintended disturbance. Implicit in the above
is the assumption that any disturbance left, as far as it influences the outcome left, 
is not related to the coin toss nor to the potential outcomes right, and vice versa.

\begin{figure}
\begin{equation*}
\boxed{
\setlength{\unitlength}{1cm}
\begin{picture}(4,4)(0.15,5.1)
\put(1.36,8.20){\line(1,0){1.52}}
\put(1.08,7.96){\line(0,-1){1.63}}
\put(3.22,7.99){\line(0,-1){1.70}}
\put(1.39,6.17){\line(1,0){1.56}}
\put(1.96,5.66){\rmfamily =?}
\put(3.39,7.08){\rmfamily =?}
\put(0.474,7.08){\rmfamily =?}
\put(3.01,5.89){$Y_2$}
\put(0.745,5.89){$X_1$}
\put(0.745,8.16){$Y_1$}
\put(3.01,8.16){$X_2$}
\put(1.93,8.30){\rmfamily =?}
\end{picture}
}
\end{equation*}
\caption{The number of equalities is even.}
\label{f.2}
\end{figure}
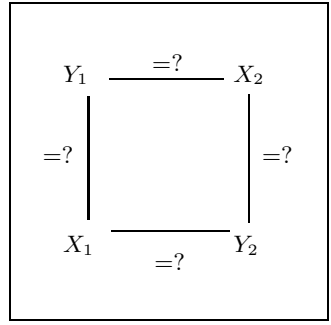

We now procede to prove Bell's inequality, and from this, Bell's theorem.
The first step is to note a logical fact: arrange the four binary variables
$X_1$, $Y_1$, $X_2$, $Y_2$ at the corners of a square, in the sequence just
given. Each of the four sides of the square 
connects one of the $X_i$ with one of the $Y_j$. Now for
each pair $X_i,Y_j$ ask the question: do these variables take the same value, or
are they different; see Figure~\ref{f.2}. 
One sees immediately that any three equalities imply the fourth; 
and also that any three inequalities imply the fourth. 
It follows that \emph{the number of equalities always equals $0$, $2$ or $4$}. 
For an algebraic proof of this fact, note that the value of $X_iY_j$ 
encodes the equality or inequality of the variables $X_i$ and $Y_j$, 
while $(X_1Y_2)=(X_1Y_1)(X_2Y_1)(X_2Y_2)$.

Next, define $\idty\{\dots\}$ as the indicator variable of a specified event;
that is to say, the indicator variable is the variable 
which takes the value  $1$ if the event happens, 
$0$ if not. Consider 
\begin{equation}\label{e:delta}
\idty\{X_1=Y_2\}-\idty\{X_1=Y_1\}-\idty\{X_2=Y_1\}-\idty\{X_2=Y_2\}~=~\Delta.
\end{equation}
By what we have just said, $\Delta$ can only take on the values $0$ and $-2$,
hence its expected value is not greater than $0$. Now, the expected value of
a linear combination of random variables equals the same linear combination of
the expected values of each variable separately. Moreover, the expected value of
an indicator variable (which only takes the values $0$ and $1$) is equal to the 
\emph{probability} of the value $1$, thus equals the probability of the event in question.
Therefore
\begin{equation}\label{e:Edelta}
\mathrm \Pr\{X_1=Y_2\}
-\Pr\{X_1=Y_1\}-\Pr\{X_2=Y_1\}-\Pr\{X_2=Y_2\}~=~\mathrm E(\Delta)~\le~0.
\end{equation}
Now consider the conditional probability that the outcomes left and right are equal, given any 
pair of measurement settings, $\Pr\{X=Y\mid AB=ij\}$. By \emph{local realism}, this
equals $\Pr\{X_i=Y_j\mid AB=ij\}$. But by \emph{freedom} this conditional probability is
the same as the unconditional probability $\Pr\{X_i=Y_j\}$.
Therefore we obtain Bell's inequality:
\begin{equation}\label{e:bellinequ}
\begin{aligned}
\Pr\{X=Y&\mid AB=12\}\\
&-\Pr\{X=Y\mid AB=11\}-\Pr\{X=Y\mid AB=21\}-\Pr\{X=Y\mid AB=22\}~\le~0.
\end{aligned}
\end{equation}
But quantum mechanics makes the prediction, for the
familiar choice of state and polarizer settings, that the expression on the left hand
side of this inequality equals $\sqrt 2 -1 \gg 0$. Hence Bell's theorem: 
if quantum mechanics holds, local realism is untenable.

\section{Hess and Philipp's Letter}
Having presented Bell's theorem in this compact form we proceed to discuss 
Hess and Philipp's difficulties with it. First of all, the only statistical independence we
needed was between the chosen polarizer settings on the one hand
and the physical system of polarizers and source on the other. Under a local
realistic model, any kinds of dependencies between hidden variables in any of
the locations source, polarizer left, polarizer right is allowed.

Secondly, we did not mention time in our derivation at all because it was 
completely irrelevant. Our derivation concerned each time point, or perhaps
better said, each time interval, within which one trial of the experiment is carried
out. We did not compare \emph{actual} outcomes under different settings at
\emph{different} times, but \emph{potential} outcomes under different
settings at the \emph{same} time. Therefore, the  argument in \cite{hp-epl} 
formulas (8)--(10) is completely besides the point.

In fact, as a thought experiment, consider repeating the measurement procedure
just described, not at a sequence of successive time intervals at the same locations,
but in a million laboratories all over the galaxy, simultaneously.
The prediction of local realism is that when we collect the one million sets of
observed quadruples $(A,B,X,Y)$ together and compute four relative frequencies
estimating the four conditional probabilities $\Pr\{X=Y\mid AB=ij\}$, they will satisfy (up
to statistical error) Bell's inequality. It is of no importance that the distribution of
hidden variables at different locations of the experiment might vary.
Moreover one of us in an earlier publication \cite{gill}
has shown using the probabilistic theory of \emph{martingales}
how the above arguments can be sharpened and applied to a time-sequence 
of measurements at the same locations, as long as new random coin tosses,
independent of the past, are used to select new settings for each trial.
The result is an exponential Chebyshev-like probability inequality concerning 
the size of experimentally observed deviations from Bell's inequality: 
neither time dependence nor time variation make extreme deviations 
any more likely than with simultaneous, independent trials.

Do Hess and Philipp then provide any arguments against our assumptions?
Hess and Philip allow in their formalism that the outcomes
on one side are related to the settings on the other side.  This they can 
only do by denying freedom or locality (or both).

Concerning freedom, their thesis would have to be that because of systematic
long-time periodicities in the various component physical systems concerned,
the outcomes of a complex series of events involving a card shuffle, a coin toss
and the free will of an experimenter at one location are interdependent and 
correlated with the potential outcome of a certain polarization measurement at 
a distant location.

Concerning locality (which of course is precisely what they want to respect),
there is an even more fantastic possibility, 
connected to what we called \emph{control} above. 
When we select a ``1'' or a ``2'' on
a measurement device, by pressing the appropriate button, we have 
supposed that \emph{only} our choice has an impact on the subsequent physics
at this location. However it is clear that at the same time we will be introducing
an uncontrolled disturbance alongside the intended binary input. 
In conceptual repetitions of the experiment, the length of time our finger presses
the button, how hard we press the button, when precisely we do it, and so on,
will vary, and each time a different though small disturbance is introduced into
the measurement device. Could it be that this disturbance actually carries with
it information about the setting being chosen in the far wing of the experiment?
Well, perhaps there is a physics in which the outcomes of coin tosses,
polarization measurements, and whether or not a physicist gets funding for
his experiment, are determined long in advance of the events, 
and are encoded in minute variations in timing and pressure, 
so that the setting being generated by Bob is in fact already ``known'' at 
Alice's location and is unwittingly introduced by her
into her apparatus along with her own coin toss (spooky, indeed).
Then the outcome left, under different hypothetical settings right, 
could differ, and our locality assumption would fail, even though the statistical
independence between the ``nominal'' instructions $A,B$ 
and the ``hidden variables'' $X_{ij},Y_{ij}$ still held.

Finally, Hess and Philipp \cite{hp-quant-ph,hp-pnas2} 
construct an elaborate hidden variables
model for the EPR correlations.  They claim that it ``does not violate Einstein 
separability''. Their argument for this claim consists merely of
a verification that the marginal probability distribution of the local variables at
one measurement station does not depend on the setting at the other station.
However this is a necessary condition, not a sufficient condition, for being
able to generate local variables with this joint probability distribution at the two locations,
without communicating in some way the setting from one wing of the experiment
to the other. In fact, we already have that the marginal distribution of the
\emph{outcome} left does not depend on the setting right. If their argumentation
were correct, one could already conclude from the probability distribution of the 
observable data (both predicted by quantum mechanics and observed in the laboratory)
that there is no problem with locality, and their explicit construction of a 
hidden variable model would be a waste of time.

\section{Conclusion}
In conclusion, Hess and Philipp have not succeeded in demolishing Bell.
They ignore the freedom of the experimenter to choose either of
two settings at the same time.  Their hidden-variables model is not local:
it requires both settings to be known at both locations in advance.
The fact that actually only one setting is in force at the time of one measurement 
is irrelevant to the proof of Bell's theorem. 
The issue of possible time-like dependence and variation is a red herring.

\acknowledgements
M.\.{Z}.\ acknowledges KBN grant No.\ 5 P03B 088 20, and the
Austrian-Polish programme
\emph{Quantum Communication and Quantum Information (2002--2003)}.

\end{document}